\documentclass[amsmath,amssymb,prl,superscriptaddress,preprint]{revtex4-1}

\usepackage{amsmath,amssymb,amsfonts,mathrsfs}
\usepackage{graphicx,soul}
\usepackage{longtable}
\usepackage[ansinew]{inputenc}
\usepackage{color,ulem}
\usepackage{SIunits}
\usepackage{bbold}
\newcommand{\fcc}{Fe$_{\rm{fcc}}$}
\newcommand{\hcp}{Fe$_{\rm{hcp}}$}

\newcommand {\aak}[1]{{\color{black} #1}}
\newcommand{\mub}{\mu_{\mathrm{B}}}

\newcommand{\ud}{\mathrm{d}}

\sethlcolor{yellow}
\begin{document}


\title{Spin excitations of individual Fe atoms on Pt(111): impact of the site-dependent giant substrate polarization}

\author{A. A. Khajetoorians}
\email[Corresponding author: ]{akhajeto@physnet.uni-hamburg.de}
\affiliation{Institute for Applied Physics, Universit\"{a}t Hamburg, D-20355 Hamburg, Germany}

\author{T. Schlenk}
\affiliation{Institute for Applied Physics, Universit\"{a}t Hamburg, D-20355 Hamburg, Germany}

\author{B. Schweflinghaus}
\affiliation{Peter Gr\"{u}nberg Institut and Institute for Advanced Simulation, Forschungszentrum J\"{u}lich \& JARA, J\"{u}lich, Germany}

\author{M. dos Santos Dias}
\affiliation{Peter Gr\"{u}nberg Institut and Institute for Advanced Simulation, Forschungszentrum J\"{u}lich \& JARA, J\"{u}lich, Germany}

\author{M. Steinbrecher}
\affiliation{Institute for Applied Physics, Universit\"{a}t Hamburg, D-20355 Hamburg, Germany}

\author{M. Bouhassoune}
\affiliation{Peter Gr\"{u}nberg Institut and Institute for Advanced Simulation, Forschungszentrum J\"{u}lich \& JARA, J\"{u}lich, Germany}

\author{S. Lounis}
\email{s.lounis@fz-juelich.de}
\affiliation{Peter Gr\"{u}nberg Institut and Institute for Advanced Simulation, Forschungszentrum J\"{u}lich \& JARA, J\"{u}lich, Germany}

\author{J. Wiebe}
 \email{jwiebe@physnet.uni-hamburg.de}
\affiliation{Institute for Applied Physics, Universit\"{a}t Hamburg, D-20355 Hamburg, Germany}

\author{R. Wiesendanger}
\affiliation{Institute for Applied Physics, Universit\"{a}t Hamburg, D-20355 Hamburg, Germany}

\date{\today}

\begin{abstract}
\noindent We demonstrate using inelastic scanning tunneling spectroscopy (ISTS) and
simulations based on density functional theory that the
amplitude and sign of the magnetic anisotropy energy for a single Fe atom adsorbed onto the
Pt(111) surface can be manipulated by modifying the adatom binding site. Since the magnitude of the
measured anisotropy is remarkably small, up to \aak{an order of magnitude smaller} than previously reported, electron-hole
excitations are weak and thus the spin-excitation
exhibits long lived precessional lifetimes compared to the values found for the same adatom on noble metal surfaces.
\end{abstract}



\maketitle%

The ability to encode magnetic information in the limit of single atoms deposited on surfaces (adatoms) relies crucially on understanding and controlling the magnetic anisotropy energy (MAE) and the underlying magnetization dynamics.
The observation of giant MAE of Co adatoms on the Pt(111) surface~\cite{Gambardella2003} has spurred many experimental and theoretical investigations of this property in different nanosystems, towards the final goal of stabilizing a single magnetic adatom.
Two techniques have emerged over the last decade which allow for single atomic spin detection, namely inelastic scanning tunneling spectroscopy (ISTS)~\cite{Heinrich2004,Hirjibehedin2007,Balashov2009,Khajetoorians2010,Khajetoorians2011} and spin-resolved STS~\cite{Meier2008,Khajetoorians2011,Khajetoorians2012}.
While hysteresis has yet to be found for an isolated single adatom on a non-magnetic surface, it has recently been shown by these techniques that artificially constructed ensembles of a few magnetic atoms show evidence of stability as a result of either ferromagnetic or antiferromagnetic exchange interactions within the ensemble~\cite{Loth2012,Khajetoorians2013}.
In these examples the substrate is paramount for establishing the magnetic properties of the ensemble and can dramatically affect the spin dynamics. Ultimately, tailoring the magnetic properties on such length scales requires a proper description of the strong hybridization between the adatoms and the surface, and how this affects the static and dynamic properties of the magnetic moments.

It remains an open question how to appropriately describe the magnetization dynamics of atomic spins placed on non-magnetic surfaces, as hybridization can dramatically alter the magnetism of the adatom. A simple approximation is to describe the impurity as a molecular magnet, namely to treat the magnetic moment as a quantized spin, and approximate the crystal field produced by the substrate in terms of powers of spin operators~\cite{Gatteschi2006}.
While these approaches describe transition metal adatoms on substrates where the atomic 3$d$ states are well localized~\cite{Hirjibehedin2007,Loth2010,Khajetoorians2010}, they fail to capture the importance of itinerant effects, like electron-hole excitations, which arise when the magnetic moment is strongly coupled to conduction electrons, as on a metallic surface~\cite{Mills1967,Lounis2010}.
As we have previously shown, the itinerant character of metallic surfaces must be considered in order to account for the measured precessional lifetimes and the $g$-shifts of Fe adatoms~\cite{Khajetoorians2011,Chilian2011}.

We report here on a surprising behavior: by monitoring the magnetic excitations of individual atoms with ISTS, we show that Fe adatoms on Pt(111) exhibit \aak{a relatively} low MAE and long precessional lifetime. Moreover, these properties are strongly dependent on which hollow site the adatom occupies. These findings are in stark contrast to those of Ref.~\cite{Balashov2009}: inelastic excitations, seen in the absence of a magnetic field, with characteristic energies of 10 meV and 6 meV for Co and Fe respectively, were interpreted as magnetic excitations with extremely short precessional lifetimes.
After carefully reexamining the case of Fe adatoms, we conclude that the MAE is \aak{an order of magnitude} weaker and the precessional lifetimes are \aak{up to} two orders of magnitude longer than originally reported.
Magnetic field dependent measurements confirm these findings and reveal that the type of binding site can totally reorient the preferred orientation of the magnetic moment (parallel/perpendicular to the surface), and affect the strength of the MAE ($E_a$), the precessional lifetime ($\tau$), and $g$-factor, as demonstrated by atomic manipulation.
We recapture these experimental observations utilizing first-principles approaches based upon time-dependent density functional theory (TD--DFT), from which we compute the MAE and magnetic excitations, and compare them with \aak{effective spin Hamiltonian model} calculations of the magnetic excitation spectra.  We show that the binding site dependence of the giant Pt polarization cloud created by the Fe adatoms is crucial for \aak{describing the MAE and the} spin dynamics, revealing the itinerant nature of the system.

Scanning tunneling spectroscopy (STS) was performed in a home-built UHV STM facility at a base temperature of  $T=0.3$ K and in magnetic fields, $B$, up to $12$ T applied perpendicular to the sample surface~\cite{Wiebe2004}.
The STM tip was etched from tungsten wire and \textit{in-situ} flashed to remove residual contaminants.
The Pt(111) surface was cleaned \textit{in-situ} by repeated cycles of Ar$^+$ sputtering and annealing to $T = 740^{\circ}$C, with a final flash at $T= 1000^{\circ}$C.
Subsequently, the clean surface was cooled to $T \approx 4$ K and exposed to Fe resulting in a distribution of single Fe atoms on the surface residing at two surface hollow sites (fcc, hcp)~\cite{supp}.
The differential conductance ($\text{d}I/\text{d}V$) was recorded with the feedback off via a lock-in technique with a modulation voltage of $V_{\rm mod}=40-200$ $\mu$V and modulation frequency $f_{\rm mod}=4.1$ kHz.

\begin{figure}[t!]
\includegraphics[width=0.8\columnwidth]{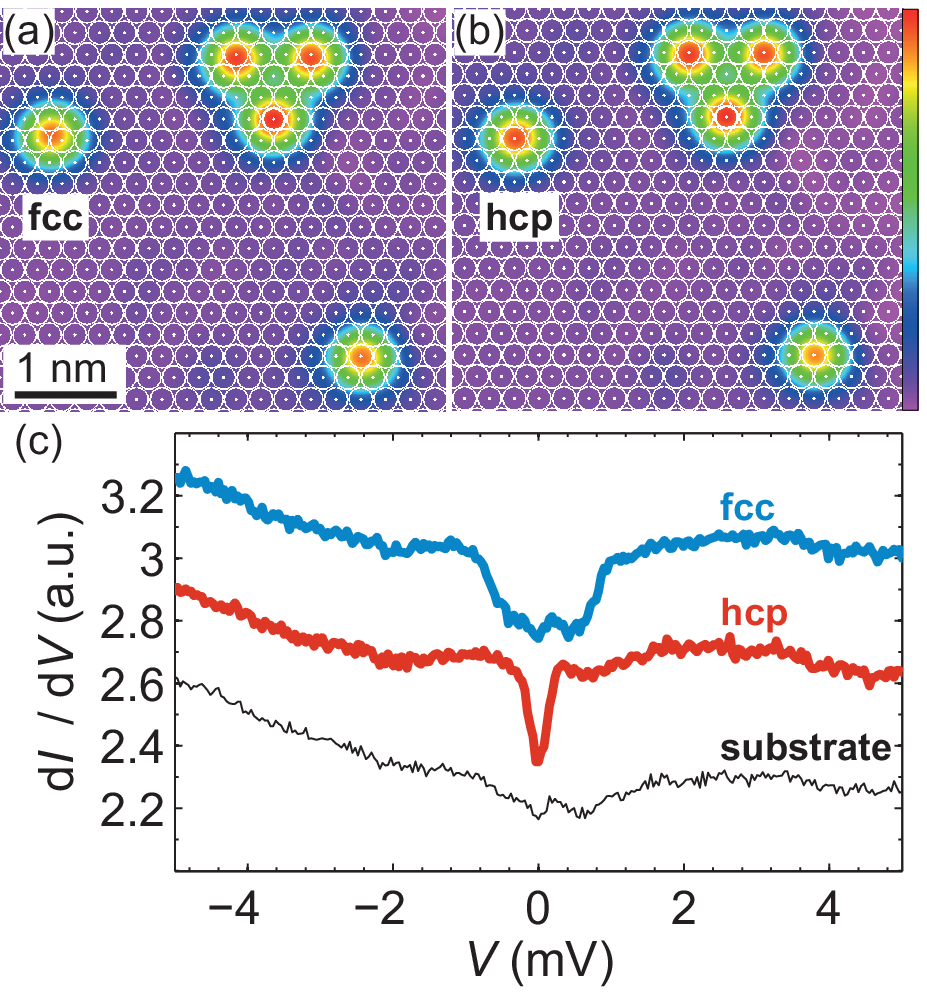}
\caption{\label{topo} STM constant-current images (a) before and (b) after manipulating the top left Fe adatom from an fcc to an hcp hollow site on Pt(111).
The center of the drawn black atomic lattice corresponds to one of two possible hollow sites. ($V_{\rm{S}}=6$ mV, $I_{\rm{t}}= 500$ pA, $T =0.3$ K; manipulation parameters: $V_{\rm{S}}=2$ mV, $I_{\rm{t}}=50$ nA).
The colorscale represents $\Delta z = 0.12$ nm.
(c) ISTS of an Fe adatom at an hcp site (red) and an fcc site (blue) as compared to the background spectrum on the Pt(111) substrate (black).
Each spectrum is vertically offset for clarity (stabilization: $V_{\rm{S}}=6$ mV, $I_{\rm{t}}=3$ nA, $V_{\rm{mod}}=40$ $\mu$V, $T =0.3$ K)}
\end{figure}

Fig.~\ref{topo}(a-b) illustrates atomic manipulation~\cite{Eigler1990,Khajetoorians2012,supp} of an Fe adatom residing on the Pt(111) surface induced by the STM tip between an fcc hollow site (\fcc) to an hcp hollow site (\hcp).
STS recorded on top of both \fcc~and \hcp~(Fig.~\ref{topo}(c)), before and after manipulation, exhibits strong step-like features symmetric to $E_{\rm{F}}$ below $\left|V_{\rm{S}}\right| < 1$ meV for each binding site.
These steps are characterized by their position ($E$), width ($W$) and intensity ($\mathscr{I}$).
\hcp~shows a stronger excitation intensity and a narrower width as compared to \fcc~at $B=0$ T.
STS done on many other Fe adatoms display the same behavior.
The step intensities are typically $\mathscr{I}_{\rm{fcc}} \approx 8\%$ and $\mathscr{I}_{\rm{hcp}} \approx 12\%$.
Such features can be identified as a tunneling-induced excitation of the adatom, when compared to the substrate~\cite{Stipe1998}.
Both types of spectra can be reproduced by manipulating the same atom between different binding sites, anywhere on the clean surface, demonstrating that the $E$, $W$, and $\mathscr{I}$ are binding site dependent.

\begin{figure}[t!]
\includegraphics[width=0.8\columnwidth]{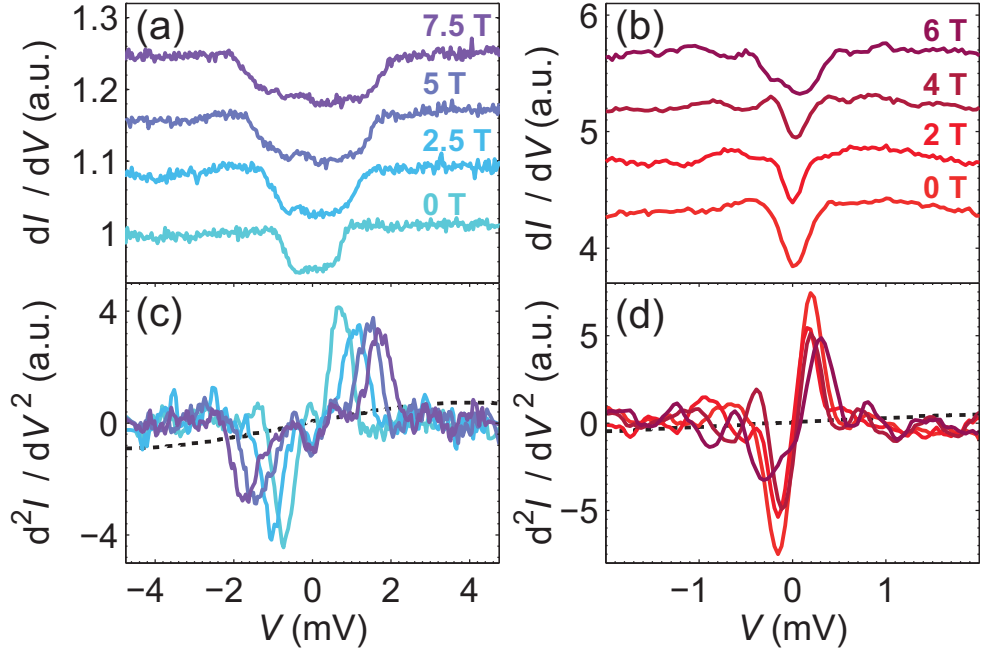}
\caption{\label{field_dep} Magnetic field dependent ISTS ($\rm{d}\it{I}/\rm{d}\it{V}$ and numerical $\rm{d}^{2}\it{I}/\rm{d}\it{V}\rm{^{2}}$) of an Fe adatom on an
fcc site (a-c, normalized to the substrate) and hcp site (b-d, unnormalized).
The spectra in (a) and (b) are offset for clarity.
. The dashed line indicates the previously reported excitation spectra for comparison~\cite{Balashov2009}.
(stabilization: $V_{\rm{S}}=6$ mV, $I_{\rm{t}}=3$ nA, $V_{\rm{mod}}=40$ $\mu$V, $T =0.3$ K)}
\end{figure}

\begin{figure}[t!]
\includegraphics[width=0.8\columnwidth]{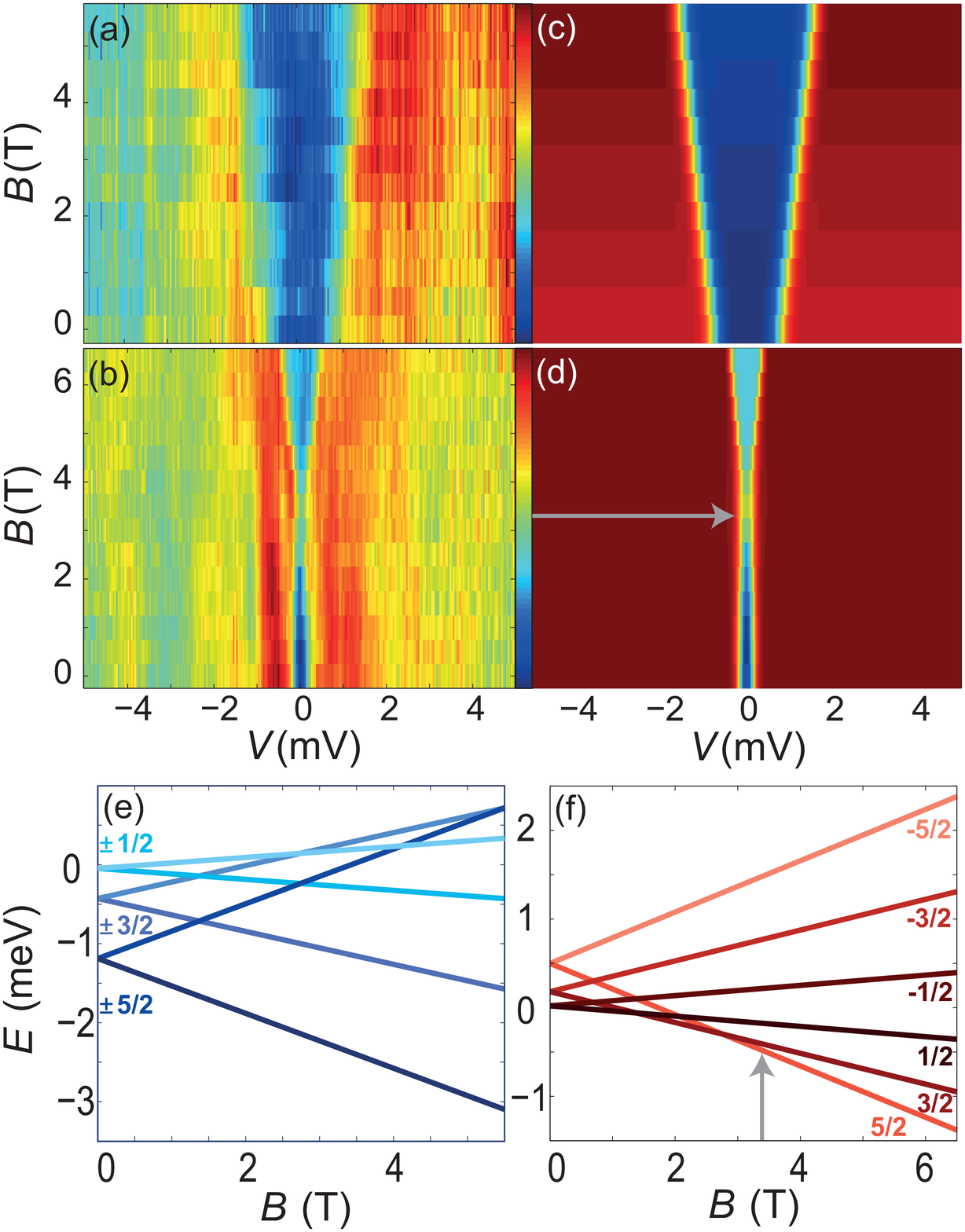}
\caption{\label{colorplot} Magnetic field dependence of the measured ISTS intensity for
(a) \fcc~(b) \hcp~and the simulated ISTS intensity based on the \aak{effective spin Hamiltonian} (see text) assuming:
(c) $D_{\rm{fcc}} = -0.19$ meV, $J_{\rm{fcc}} = 5/2$, $g_{\rm{fcc}}=2.4$, $u_{\rm{fcc}} = 2.3$, $T_{\rm{eff}}^{\rm{fcc}}=2$ K
(d) $D_{\rm{hcp}} = 0.08$ meV, $J_{\rm{hcp}} = 5/2$, $g_{\rm{hcp}}=2$, $u_{\rm{hcp}} = 2.3$, $T_{\rm{eff}}^{\rm{hcp}} =0.8$ K.
Level diagrams for (e) \fcc~and (f) \hcp. The eigenvalues $M_J$ of $\hat{J}_z$ are indicated by numbers.
(stabilization: $V_{\rm{S}}=6$ mV, $I_{\rm{t}}=3$ nA, $V_{\rm{mod}}=40$ $\mu$V, $T =0.3$ K)}
\end{figure}

To confirm that we measure inelastic magnetic excitations, we apply a magnetic field~\cite{Heinrich2004} and follow the behavior of the $\ud I/\ud V$ spectra and their numerical derivatives $\ud^{2}I/\ud V^{2}$ (Fig.~\ref{field_dep}(a-d)).
The finite zero-field excitation energy ($E_{\rm{gap}}$), is typically $E_{\rm{gap}}^{\rm{fcc}} \approx 0.75$ meV and $E_{\rm{gap}}^{\rm{hcp}} \approx 0.19$ meV.
For \fcc, $E$ shows a linear increase as the magnetic field increases (Fig.~\ref{colorplot}(a)), like seen for Fe atoms on both Cu(111) and Ag(111)~\cite{Khajetoorians2011,Chilian2011}. On the other hand, \hcp~shows an interesting non-linear behavior in $E$, $W$, and $\mathscr{I}$ as the field is increased (Fig.~\ref{colorplot}(b)).
For magnetic fields in the range of $B = 0 - 3.5$ T, there is a plateau-like behavior, namely $E$, $W$, and $\mathscr{I}$ only change slightly.
For $B > 3.5$ T, the magnetic excitation shows a linearly increasing trend in $E$, $\mathscr{I}$, $W$ similar to \fcc.
In the following, these disparate trends are interpreted as consequences of an out-of-plane MAE for \fcc~and an easy plane MAE for \hcp.

To analyze the connection between the MAE and the binding site, we performed DFT calculations with the Korringa--Kohn--Rostoker Green function method (KKR--GF) in a real-space approach~\cite{Papanikolau2002,Bauer2013}.
Pt(111) is notoriously challenging because of its high magnetic polarizability~\cite{Sipr2010,Meier2011}, owing to an extended polarization cloud which surrounds the magnetic adatom, like seen for \aak{Pd~\cite{Nieuwenhuys1975,Oswald1986,Hafner2010}}. In this light, we carefully checked all calculations. For computational details see~\cite{supp}.
The computed spin moments are $3.40\;\mub$ ($4.42\;\mub$) for \fcc~and $3.42\;\mub$ ($4.57\;\mub$) for \hcp, where the values refer to the adatom (whole 3D cluster --- 62 Pt atoms), respectively.
The orbital moments are for \fcc~ $0.11\;\mub$ ($0.23\;\mub$); \hcp~$0.08\;\mub$ ($0.22\;\mub$).
The MAE yields $E_{a}^{\rm{fcc}} = -2.05$ meV (out-of-plane) and $E_{a}^{\rm{hcp}} = +0.50$ meV (easy plane).
\aak{Here, it was crucial to include a large number of substrate atoms in order to converge the calculation~\cite{supp}.  For a small cluster with 10-12 Pt atoms, calculations of both \hcp~and \fcc~yield an out-of-plane easy axis with values for the MAE in-line with those calculations based on a supercell KKR-GF method~\cite{Balashov2009}. However, only after including more than 60 Pt atoms, the calculated MAE finally converges and reveals a reorientation of the MAE of \hcp~into the easy-plane configuration. This shows that the spin polarization of the substrate generated by each Fe adatom type effectively reduces the total MAE, as similarly discussed in Ref.~\cite{Hafner2010}.}

Fig.~\ref{colorplot}(a-b) show results of magnetic field-dependent spectra with high energy resolution, at smaller field steps $\Delta B =0.5$~T. A subset of this data was already shown in Fig.~\ref{field_dep} for clarity.
Following~\cite{Hirjibehedin2007,Loth2010,Khajetoorians2010}, \aak{an effective spin Hamiltonian} model is used for phenomenological analysis: $\hat{\mathcal H}_{J} = D\hat{J}_{z}^{2} + g\mu_{\rm{B}}B\hat{J}_z$~\cite{Gatteschi2006,Dai2008,supp}. This is the sum of the anisotropy energy and the Zeeman energy.
The model parameters are eigenvalue $J$, the anisotropy constant $D$ (negative for out-of-plane easy axis and positive for easy plane) and the $g$-factor. $B$ is the applied magnetic field which is out-of-plane here.
The theoretical excitation spectra shown in Fig.~\ref{colorplot}(c-d) are derived by considering an interaction $\hat{s} \cdot \hat{J} + u\hat{\mathbb{1}}$ between the tunneling electron and the impurity~\cite{Lorente2012,Rossier2009,Fransson2009,Chilian2011a}. While the first term describes the exchange interaction between the tunneling electron spin $\hat{s}$ and the atomic spin $\hat{J}$, $u$ quantifies the strength of elastic tunneling. As the hybridization of the moment with the substrate is strong, the assumption of an isolated effective spin is not justified.  Therefore, we mimic the effect of the substrate electrons by introducing an artificial broadening of the excitation steps using an effective temperature $T_{\rm{eff}}$ to fit the experimental $W$, where $T_{\rm{eff}}^{\rm{fcc}} = 2$ K and $T_{\rm{eff}}^{\rm{hcp}} = 0.8$ K. \aak{The value of $J$ was chosen to be closest to the DFT calculated total magnetic moments of the whole cluster which includes the surrounding substrate, namely $J = 5/2$ for both \fcc~and \hcp.}
However, the qualitative behavior is the same for other values of $J$, as the sign of $D$ determines the phenomenology.

The results of modeling the data in Fig.~\ref{colorplot}(a-b) are shown in Fig.~\ref{colorplot}(c-d).
Taking $D_{\rm{fcc}} = -0.19$ meV, \fcc~is understood to be always in an out-of-plane (maximum $M_J$) ground state, as the excitation energy increases linearly with $B$. For $D_{\rm{hcp}} = 0.08$ meV, \hcp~has an in-plane (minimum $M_J$) ground state when $B = 0$. The plateau region corresponds to the eventual transition of the ground state to out-of-plane (increasing $M_J$).  Once this is reached, at the indicated crossing point (gray arrow), the same linear behavior at higher fields is observed like for \fcc.  \aak{It is important to note that, in addition to the spin excitation, we cannot rule out a Kondo effect masked below the spin excitation for \hcp.  However, the Kondo temperature is most likely below our measurement temperature~\cite{supp} and is neglected since we recapture the measurement in the modeling without considering a significant Kondo effect.} To compare the modeled spectra and the values of $D$ to the DFT calculated values
the magnetic anisotropy energy $E_a$ and the model anisotropy parameter $D$ are connected by the correspondence principle: $D(J) = E_a/J(J+1)$. From the DFT calculations, we extract \aak{the values}, $D_{\rm{fcc}}(5/2) = -0.23$ meV and $D_{\rm{hcp}}(5/2) = 0.06$ meV, which are consistent with the experimentally determined model parameters.
Itinerant effects such as \aak{the broadened linewidth,} the observed shift in $g$ for \fcc,~and the field dependence of the linewidth are beyond the scope of the model and will be discussed below in the context of TD--DFT calculations of the dynamical magnetic susceptibility.

\begin{figure}[t!]
\includegraphics[width=0.8\columnwidth]{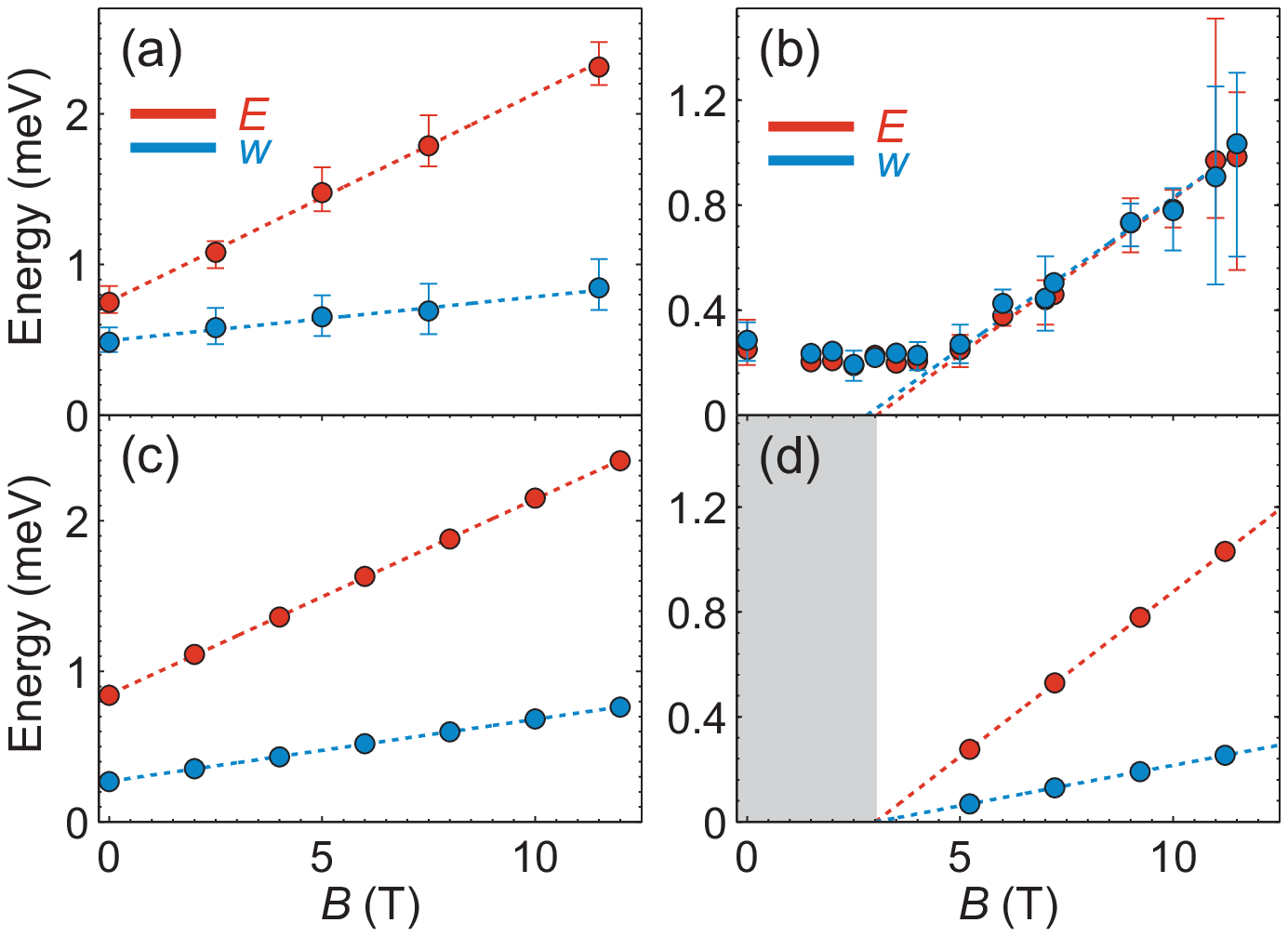}
\caption{\label{plots} The experimental excitation energy ($E$: red) and FWHM ($W$: blue) of the linewidth for (a) \fcc, (b) \hcp,
and the calculated excitation energy \aak{(red) and FWHM (blue) for (c) \fcc~and (d) \hcp~extracted from the $\mathrm{Im}(\chi)$} (the gray rectangle represents the magnetic field range where the \hcp~ground state is not the maximum $M_J$ state).
Dashed lines represent the linear fit and the error bars represent the largest range of measured values.
The effective energy resolution in (a-b) was $\Delta E \approx 0.12 - 0.3$ meV.}
\end{figure}

The precessional lifetime $\tau$ and $g$-factor were extracted by measuring $E$ and $W$ (FWHM) as a function of magnetic field for many Fe atoms (Fig.~\ref{plots}(a-b)).
We extract $\tau$ at zero field by considering $\tau = \hbar/(2 W_0)$, where $W_0$ is the intrinsic linewidth~\cite{footnote1} derived from gaussian fitting the numerically derived $\ud^2 I/\ud V^2$ spectra (Fig.~\ref{field_dep}(c-d)).
The $g$-factor, where $g = \ud E/\ud(\mub B)$, was determined from a linear fit to $E(B)$ (after the plateau, in the case of \hcp).
For \fcc~an enhanced $g$-factor is measured, $g_{\rm{fcc}} = 2.4 \pm 0.1$, and $\tau_{\rm{fcc}}(B\!=\!0\,\rm{T}) = 0.70 \pm 0.12$ ps.
The $g$-factor of \hcp~was fitted for $B > 3.5$~T, yielding $g_{\rm{hcp}} = 2.0 \pm 0.15$.
The measured precessional lifetime is as large as $\tau_{\rm{hcp}}(B\!=\!0\,\rm{T}) \approx 2.5$ ps~\cite{footnote2}.

The measured values are in good agreement with the dynamical \aak{transverse} magnetic susceptibility $\chi$ computed from TD--DFT combined with the KKR--GF method~\cite{Lounis2010,Lounis2011}.
The effect of spin-orbit coupling is approximated by including an additional magnetic field which mimics $E_{\rm{gap}}$.
From the imaginary part of $\chi$, which gives the density of states for spin excitations, we extract the calculated excitation energy and width as a function of $B$, shown in Fig.~\ref{plots}(c-d)~\cite{supp}. By linear fits, we then extract $g$ and $\tau$.
We obtain $g_{\rm{fcc}} = 2.24$ and $g_{\rm{hcp}} = 2.18$, illustrating the trend that \fcc~maintains a higher $g$-value as compared to \hcp.
Inputting the experimental $E_{\rm{gap}}$ for both cases, the calculated $\tau$ is found to be larger for \hcp~(4.8 ps) than for \fcc~(1.2 ps), as experimentally observed.
As spin-orbit coupling was not included in these calculations, it is possible that it can modify the computed values of the $g$-factor and of $\tau$.
The shift in $g$ and the reduction of $\tau_{\rm{fcc}}(B\!>\!0\,\rm{T})$, $\tau_{\rm{hcp}}(B\!>\!3.5\,\rm{T})$ for increasing magnetic field result from spin-dependent scattering by conduction electrons (Stoner excitations) which damp the spin precession, as previously observed in related systems\aak{~\cite{Khajetoorians2011,Lounis2010,Lounis2011,Chilian2011}}.
Unlike Fe atoms on both the Cu(111) and Ag(111) surfaces, Fe atoms on Pt(111) show comparatively larger precessional lifetimes (due to the lower excitation energies), which decrease more weakly ($\rm{d}{\it \tau}/\rm{d}{\it B}$) in a magnetic field than in the aforementioned systems.

Previous measurements of inelastic excitations of single Fe atoms on Pt(111)~\cite{Balashov2009}, done in the absence of a magnetic field, reported only one adsorption site, unlike the two observed here, which exhibits a much smaller excitation intensity (dashed line Fig.~\ref{field_dep}(c-d)) occuring at energies $7 - 30$ times higher than the energies at which we unambiguously observe magnetic excitations.
Measurements performed as a function of temperature, $T = 0.3 - 4.3$ K~\cite{supp} do not exhibit any inelastic excitations for clean Fe adatoms, up to tunneling currents $I_{\rm{t}} \leq 30$ nA, that resemble those seen in Ref.~\onlinecite{Balashov2009}. They do reveal, however, that at $T = 4.3$ K only \fcc~ displays a clear magnetic excitation but at an energy much lower than the previously reported value. The effect of temperature simply broadens the excitation but does not shift it. Aside from the striking dependence of the magnetism on the binding site dependence, the values of $\tau$ measured here are two orders of magnitude larger than those reported in ref.~\cite{Balashov2009}.

In conclusion, we find that Fe adatoms on Pt(111) exhibit a remarkably small MAE, in stark contrast to Co atoms on Pt(111)~\cite{Gambardella2003}.
The measured values are substantially lower compared to what was previously reported~\cite{Balashov2009}, as well as compared to lighter substrates~\cite{Khajetoorians2011,Chilian2011}.
Previous XMCD measurements of Fe/Pt(111) suggested small values of the MAE~\cite{Lehnert2009}, but the site dependence and magnitude of this quantity could not be extracted.
\aak{Moreover, the surprising finding that the type of occupied hollow site can completely alter the orientation of the magnetic moment is illuminated by DFT when considering the contribution of the large polarization cloud induced in the Pt substrate. A similar binding site dependence of the MAE was previously predicted for Fe adatoms on Pd(111)~\cite{Hafner2010}.} Our measurements and calculations reveal that, while Pt(111) sustains such a large polarization cloud (we consider a radius $\approx 0.75$ nm), it also gives rise to longer lifetimes and relatively weak damping due to Stoner excitations for the Fe adatoms as compared to magnetic excitations of Fe on other noble metal surfaces~\cite{Khajetoorians2011,Chilian2011}.
This goes against what might be expected from the stronger hybridization between the $d$-states of the adatoms and the $d$-states of Pt, as compared with the $sp$-states near the Fermi energy from the Cu and Ag substrates.
Given that the lifetime of the spin precession is \aak{inversely} proportional to the excitation energy, the much smaller zero field magnetic excitation gap, controlled by the low MAE, is responsible for this behavior.
These results illustrate that the behavior of Fe/Pt(111), a typical system used for out-of-plane device technology, can dramatically change when scaled to the atomic limit.

We would like to thank H. Brune, W. Wulfhekel, A. Lichtenstein, V. Caciuc, G. Bihlmayer, and S. Bl\"{u}gel
for fruitful discussions. A. A. K., T. S., M. S., J. W. and R. W. acknowledge
funding from SFB668-A1 and GrK1286 of the DFG and from the ERC Advanced Grant ``FURORE.'' \aak{A.A.K. also acknowledges Project no.~KH324/1-1 from the Emmy-Noether-Program of the DFG.} B. S., M. S. D., M. B. and S. L. acknowledge support of the HGF-YIG Programme VH-NG-717 (Functional Nanoscale Structure
and Probe Simulation Laboratory-Funsilab).

\newpage

\begin{thebibliography}{29}
\expandafter\ifx\csname natexlab\endcsname\relax\def\natexlab#1{#1}\fi
\expandafter\ifx\csname bibnamefont\endcsname\relax
  \def\bibnamefont#1{#1}\fi
\expandafter\ifx\csname bibfnamefont\endcsname\relax
  \def\bibfnamefont#1{#1}\fi
\expandafter\ifx\csname citenamefont\endcsname\relax
  \def\citenamefont#1{#1}\fi
\expandafter\ifx\csname url\endcsname\relax
  \def\url#1{\texttt{#1}}\fi
\expandafter\ifx\csname urlprefix\endcsname\relax\def\urlprefix{URL }\fi
\providecommand{\bibinfo}[2]{#2}
\providecommand{\eprint}[2][]{\url{#2}}

\bibitem[{\citenamefont{Gambardella et~al.}(2003)\citenamefont{Gambardella,
  Rusponi, Veronese, Dhesi, Grazioli, Dallmeyer, Cabria, Zeller, Dederichs,
  Kern et~al.}}]{Gambardella2003}
\bibinfo{author}{\bibfnamefont{P.}~\bibnamefont{Gambardella} {\it et al.}},
  \bibinfo{journal}{Science} \textbf{\bibinfo{volume}{300}},
  \bibinfo{pages}{1130} (\bibinfo{year}{2003}).

\bibitem[{\citenamefont{Heinrich et~al.}(2004)\citenamefont{Heinrich, Gupta,
  Lutz, and Eigler}}]{Heinrich2004}
\bibinfo{author}{\bibfnamefont{A.~J.} \bibnamefont{Heinrich} {\it et al.}},
  \bibinfo{journal}{Science} \textbf{\bibinfo{volume}{306}},
  \bibinfo{pages}{466} (\bibinfo{year}{2004}).

  \bibitem[{\citenamefont{Hirjibehedin et~al.}(2007)\citenamefont{Hirjibehedin,
  Lin, Otte, Ternes, Lutz, Jones, and Heinrich}}]{Hirjibehedin2007}
\bibinfo{author}{\bibfnamefont{C.~F.} \bibnamefont{Hirjibehedin} {\it et al.}},
  \bibinfo{journal}{Science} \textbf{\bibinfo{volume}{317}},
  \bibinfo{pages}{1199} (\bibinfo{year}{2007}).

  \bibitem[{\citenamefont{Balashov et~al.}(2009)\citenamefont{Balashov, Schuh,
  Tak\'acs, Ernst, Ostanin, Henk, Mertig, Bruno, Miyamachi, Suga
  et~al.}}]{Balashov2009}
\bibinfo{author}{\bibfnamefont{T.}~\bibnamefont{Balashov} {\it et al.}},
  \bibinfo{journal}{Phys. Rev. Lett.} \textbf{\bibinfo{volume}{102}},
  \bibinfo{pages}{257203} (\bibinfo{year}{2009}).

  \bibitem[{\citenamefont{Khajetoorians et~al.}(2010)\citenamefont{Khajetoorians,
  Chilian, Wiebe, Schuwalow, Lechermann, and Wiesendanger}}]{Khajetoorians2010}
\bibinfo{author}{\bibfnamefont{A.~A.} \bibnamefont{Khajetoorians} {\it et al.}},
  \bibinfo{journal}{Nature} \textbf{\bibinfo{volume}{467}},
  \bibinfo{pages}{1084} (\bibinfo{year}{2010}).

  \bibitem{Khajetoorians2011}
A.~A.~Khajetoorians, {\it et~al.\/}, Itinerant nature of atom-magnetization excitation by tunneling electrons, {\it Phys. Rev. Lett.} {\bf 106}, 037205 (2011).

  \bibitem[{\citenamefont{Meier et~al.}(2008)\citenamefont{Meier, Zhou, Wiebe,
  and Wiesendanger}}]{Meier2008}
\bibinfo{author}{\bibfnamefont{F.}~\bibnamefont{Meier} {\it et al.}},
  \bibinfo{journal}{Science} \textbf{\bibinfo{volume}{320}},
  \bibinfo{pages}{82} (\bibinfo{year}{2008}).

  \bibitem[{\citenamefont{Khajetoorians et~al.}(2012)\citenamefont{Khajetoorians, Wiebe,
  Chilian, Lounis, Bl\"{u}gel, and Wiesendanger}}]{Khajetoorians2012}
\bibinfo{author}{\bibfnamefont{A.~A.} \bibnamefont{Khajetoorians} {\it et al.}},
  \bibinfo{journal}{Nature Physics} \textbf{\bibinfo{volume}{497}},
  \bibinfo{pages}{497} (\bibinfo{year}{2012}).

  \bibitem[{\citenamefont{Loth et~al.}(2012)\citenamefont{Loth, Baumann, Lutz, Eigler, Heinrich}}]{Loth2012}
\bibinfo{author}{\bibfnamefont{S.} \bibnamefont{Loth} {\it et al.}},
  \bibinfo{journal}{Science} \textbf{\bibinfo{volume}{335}},
  \bibinfo{pages}{196} (\bibinfo{year}{2012}).

  \bibitem[{\citenamefont{Khajetoorians et~al.}(2013)\citenamefont{Khajetoorians, Baxevanis, H\"{u}bner, Schlenk, Krause, Wehling, Lounis, Lichtenstein, Pfannkuche, Wiebe, and Wiesendanger}}]{Khajetoorians2013}
\bibinfo{author}{\bibfnamefont{A.~A.} \bibnamefont{Khajetoorians} {\it et al.}},
  \bibinfo{journal}{Science} \textbf{\bibinfo{volume}{339}},
  \bibinfo{pages}{55} (\bibinfo{year}{2013}).

  \bibitem{Gatteschi2006}
D.~Gatteschi, R.~Sessoli, Molecular nanomagnets (Oxford Uni. Press, Oxford, ed. 1, 2006)

  \bibitem[{\citenamefont{Loth et~al.}(2010)\citenamefont{Loth, von Bergmann,
  Ternes, Otte, Lutz, and Heinrich}}]{Loth2010}
\bibinfo{author}{\bibfnamefont{S.}~\bibnamefont{Loth} {\it et al.}},
  \bibinfo{journal}{Nature~Phys.} \textbf{\bibinfo{volume}{6}},
  \bibinfo{pages}{340} (\bibinfo{year}{2010}).

  \bibitem[{\citenamefont{Mills and Lederer}(1967)}]{Mills1967}
\bibinfo{author}{\bibfnamefont{D.~L.} \bibnamefont{Mills}} \bibnamefont{and}
  \bibinfo{author}{\bibfnamefont{P.}~\bibnamefont{Lederer}},
  \bibinfo{journal}{Phys. Rev.} \textbf{\bibinfo{volume}{160}},
  \bibinfo{pages}{590} (\bibinfo{year}{1967}).

\bibitem{Lounis2010}
S.~Lounis, A.~T. Costa, R.~B. Muniz, D.~L. Mills, Dynamical Magnetic Excitations of Nanostructures from First Principles, {\it Phys. Rev. Lett.} {\bf 105}, 187205 (2010).

\bibitem{Chilian2011}
B.~Chilian, A.~A. Khajetoorians, S.~Lounis, A.~T.~Costa, D.~L.~Mills, J.~Wiebe, R.~Wiesendanger, Anomalously large $g$ factor of single atoms adsorbed on a metal substrate, {\it Phys. Rev. B} {\bf 84}, 212401 (2011).

\bibitem[{\citenamefont{Wiebe et~al.}(2004)\citenamefont{Wiebe, Wachowiak,
  Meier, Haude, Foster, Morgenstern, and Wiesendanger}}]{Wiebe2004}
\bibinfo{author}{\bibfnamefont{J.}~\bibnamefont{Wiebe} {\it et al.}},
  \bibinfo{journal}{Rev. Sci. Inst.} \textbf{\bibinfo{volume}{75}},
  \bibinfo{pages}{4871} (\bibinfo{year}{2004}).

  \bibitem{supp}
See online supplemental material.

\bibitem{Eigler1990}
D.~M.~Eigler, E.~K.~Schweizer, Positiong single atoms with a scanning tunneling microscope, {\it Nature} {\bf 344}, 524 (1990).

\bibitem[{\citenamefont{Stipe et~al.}(1998)\citenamefont{Stipe, Rezaei, and
  Ho}}]{Stipe1998}
\bibinfo{author}{\bibfnamefont{B.~C.} \bibnamefont{Stipe}},
  \bibinfo{author}{\bibfnamefont{M.~A.} \bibnamefont{Rezaei}},
  \bibnamefont{and} \bibinfo{author}{\bibfnamefont{W.}~\bibnamefont{Ho}},
  \bibinfo{journal}{Science} \textbf{\bibinfo{volume}{280}},
  \bibinfo{pages}{1732} (\bibinfo{year}{1998}).

\bibitem{Papanikolau2002}
N.~Papanikolau, R.~Zeller, P.~H.~Dederichs, {\it J. Phys.: Condens. Matter} {\bf 14}, 2799 (2002).

\bibitem{Bauer2013}
D.S.G. Bauer, Rheinisch-Westf\"aliche Technische Hochschule (RWTH), Aachen (2013).


\bibitem{Sipr2010}
O.~Sipr, S.~Bornemann, J.~Minar, H.~Ebert, {\it Phys. Rev. B} {\bf 82}, 174414 (2010).

\bibitem{Meier2011}
F.~Meier, S.~Lounis, J.~Wiebe, L.~Zhou, S.~Heers, P.~Mavropoulos, P.~H.~Dederichs, S.~Bl\"{u}gel, R.~Wiesendanger, Spin polarization of platinum (111) induced by the proximity to cobalt nanostripes,  {\it Phys. Rev. B} {\bf 83}, 075407 (2011).

\bibitem{Nieuwenhuys1975}
G.~J.~Nieuwenhuys, Magnetic behaviour of cobalt, iron and manganese dissolved in palladium,  {\it Advances in Physics} {\bf 24}, 515 (1975).

\bibitem{Oswald1986}
A.~Oswald, R.~Zeller, P.~H.~Dederichs, Giant moments in Palladium,  {\it Phys. Rev. Lett} {\bf 56}, 1419 (1986).

\bibitem{Hafner2010}
P.~Blonski, A.~Lehnert, S.~Dennler, S.~Rusponi, M.~Etzkorn, G.~Moulas, P.~Bencok, P.~Gambardella, H.~Brune, J.~Hafner, Magnetocrystalline anisotropy energy of Co and Fe adatoms on the (111) surfaces of Pd and Rh, {\it Phys. Rev. B} {\bf 81}, 104426 (2010).

\bibitem{Dai2008}
D.~Dai, H.~Xiang, M.~H.~Whangbo, Effects of spin-orbit coupling on magnetic properties of discrete and extended magnetic systems, {\it Journal of Computational Chemistry} {\bf 29}, 2187 (2008).

\bibitem{Lorente2012}
J.~P.~Gauyacq, N.~Lorente, F.~D.~Novaes, Excitation of local magnetic moments by tunneling electrons, {\it Progress in Surface Science} {\bf 87}, 73 (2012).

\bibitem{Rossier2009}
J.~Fern\'andez-Rossier, Theory of Single-Spin Inelastic Tunneling Spectroscopy, {\it Phys. Rev. Lett.} {\bf 102}, 256802 (2009).

\bibitem{Fransson2009}
J.~Fransson, Spin inelastic electron tunneling spectroscopy on local spin adsorbed on surface, {\it Nano. Lett.} {\bf 9}, 2414-2417 (2009).

\bibitem{Chilian2011a}
B.~Chilian, A.~A.~Khajetoorians, J.~Wiebe, R.~Wiesendanger, Experimental variation and theoretical analysis of the inelastic contribution to atomic spin excitation spectroscopy, {\it Phys. Rev. B} {\bf 83}, 195431 (2011).

\bibitem{footnote1}
$W = \sqrt{(1.7V_{\rm{mod}})^2 + (5.4k_{\rm{B}}T)^2 + (W_0)^2}$

\bibitem{footnote2}
Values up to $E_{\rm{gap}}^{\rm{hcp}}  \approx 0.3$ meV and lifetimes down to $\tau_{\rm{hcp}} \approx 1.2$ ps are measured because of energy resolution broadening resulting from an increased $V_{\rm{mod}}$ utilized during particular measurements.

\bibitem{Lounis2011}
S.~Lounis, A.~T.~Costa, R.~B.~Muniz, D.~L.~Mills, Theory of local dynamical magnetic susceptibilities from the Korringa-Kohn-Rostoker Green function method, {\it Phys. Rev. B} {\bf 83}, 035109 (2011).

\bibitem{Lehnert2009}
A.~Lehnert, Swiss Federal Institute of Technology, Lausanne (2009).

\end{thebibliography}
\end{document}